\documentstyle[floats,prd,aps,epsfig,eqsecnum,12pt]{revtex}
\makeatletter
\newbox\tempboxa
\newdimen\captionboxsubcount 
\def\capsize#1{\captionboxsubcount=#1pt}
\newdimen\captionboxsub
\captionboxsub=\hsize \advance\captionboxsub by -\captionboxsubcount
\advance\captionboxsub by -\captionboxsubcount
\long\def\@makecaption#1#2{
 \setbox\@tempboxa\hbox{#1 #2}
 \ifdim \wd\@tempboxa >\captionboxsub 
\rightskip=\captionboxsubcount \leftskip=\captionboxsubcount #1 #2 
\else \hbox to\hsize{\hfil\box\@tempboxa\hfil} 
 \fi}
\makeatother
\capsize{30}

\begin{document}
\bibliographystyle{unsrt}
\begin{titlepage}
\begin{flushright}
\begin{minipage}{5cm}
\begin{flushleft}
\small
\baselineskip = 13pt
SU-4240-719\\
hep-ph/0004105 \\
\end{flushleft}
\end{minipage}
\end{flushright}
\begin{center}
\Large\bf
Complementary Ansatz for the neutrino mass matrix
\end{center}
\vfil
\footnotesep = 12pt
\begin{center}
\large
Deirdre {\sc Black} 
\footnote{Electronic address: {\tt black@physics.syr.edu}}
\quad\quad Amir H. {\sc Fariborz}\footnote{Electronic  
address: {\tt amir@suhep.phy.syr.edu}}\\
\vskip 0.5cm
Salah {\sc Nasri}\footnote{
Electronic address : {\tt snasri@suhep.phy.syr.edu}} 
\quad\quad
Joseph {\sc Schechter}\footnote{
Electronic address : {\tt schechte@suhep.phy.syr.edu}}\\
{\it  Department of Physics, Syracuse University, 
Syracuse, NY 13244-1130, USA.} \\
\vskip 0.5cm
\end{center}
\vfill
\begin{center}
\bf
Abstract
\end{center}
\begin{abstract}
We propose a simple Ansatz for the three generation neutrino mass matrix $M_\nu$
which is motivated from an SO(10) grand unified theory. The Ansatz can be
combined with information from neutrino oscillation experiments and bounds
on neutrinoless double beta decay to determine the neutrino masses
themselves and to reconstruct, with some assumptions, the matrix $ M_\nu $
.

\baselineskip = 17pt
\end{abstract}
\begin{flushleft}
\footnotesize
 
\end{flushleft}
\vfill
\end{titlepage}
\setcounter{footnote}{0}
\section{Introduction}
In the last few years there has been another wave of excitement regarding
the question of neutrino masses. This is largely due to the many new
experiments testing neutrino oscillations, most notably the positive
indications obtained by Super Kamiokande on atmospheric neutrino
oscillations \cite{SK}.  Similar indications come from other experiments
\cite{Soudan,Macro,Kamiokandeatmos,IMB}. The solar neutrino experiments have for many years
 provided independent
evidence for neutrino oscillations \cite{Sage,Gallex,Homestake,Superkamiokandesolar,Kamiokandesolar}. Accelerator and reactor experiments
have also played an important role. They have furnished strict bounds on
neutrino oscillation parameters \cite{Chooz,Bugey,Karmen,Nomad,Chorus,LSND}. In the case of
the LSND experiment \cite{LSND} at Los
Alamos evidence for $ \overline{\nu}_\mu \rightarrow
\overline{\nu}_e$ oscillation has been reported. See refs 
\cite{review} for recent reviews.

It is hoped that new experimental results can be used to determine the
neutrino squared mass differences and mixing angles. In turn, these may help
to infer the neutrino mass matrix. This is presumably a possible gateway
to a more fundamental theory beyond the standard model. Of course this is
a highly speculative area, and even though there are  many imaginative
proposals \cite{models}, it seems fair to say that the the true answer is
essentially unknown. In
order to make progress in this direction, it seems useful to investigate
various plausible Ansatze for the neutrino mass matrix. From this point of
view we propose the Ansatz for the 3 generation neutrino mass matrix, $
M_\nu $:
\begin{equation}
{\rm Tr}( M_\nu ) = 0 
\end{equation}
and investigate its consequences. We are considering the neutrinos to be
represented by 2-component spinors so that, in the most general situation, $
M_\nu $ is an arbitrary symmetric complex matrix.

As we will see in section II, Eq. (1.1) can be motivated from an SO(10) 
grand unified model \cite{SO(10)}, in which it may be derived with some
assumptions. Physically, Eq. (1.1) corresponds to the well known
approximate signature of grand unification that ${m(b) \over m( \tau
)}\simeq 3 $. Furthermore we will see in sections IV and V that Eq. (1.1) 
can be straightforwardly combined with experimental information to get an
idea of the neutrino masses themselves as well as the ``texture'' of $
M_\nu
$. Relevant matters of notation are discussed in section III while a
summary is presented in section VI. 

\section{Plausibility of the Ansatz} In the SO(10) grand unification model
each generation contains one light massive two component neutrino and also
a very heavy one which is ``integrated out'' according to the ``seesaw
mechanism'' \cite{see}. The effective $3\times 3$ neutrino mass matrix
takes the form:  \begin{equation} M_{\nu } = M_{L} - M_{D}^{T}M_{H}M_{D}
\end{equation} where $M_{L}$, $M_{H}$ and $M_{D}$ are respectively the
mass matrices of the light neutrinos, heavy neutrinos and heavy-light
mixing (or ``Dirac matrix''). Generally the second, seesaw, term is
considered to dominate. Here however we shall assume the first term to be
the dominant one. This is necessary for the present derivation of Eq.
(1.1) to hold. Also, a rough order of magnitude estimate for the second
term would be $ {m(\tau)^{2}\over 10^{16}{\rm GeV} } $ or about $3 \times
10^{-7}$ eV. Thus, the seesaw term could be negligible if neutrino masses
turn out to be appreciably larger than this value. Now in SO(10), Higgs
mesons belonging to the 10, 120 and 126 representations can contribute to
the fermion masses at tree level. One has \cite{MFN} for the down quark,
charged lepton and light neutrino mass matrices, \begin{eqnarray} M_{d}
&=& aS(10) + bA(120) - {1\over 3} cS(126) \nonumber \\ rM_{e} &=& aS(10) +
dA(120) + cS(126) \\ sM_{L} &=& eS(126) \nonumber \end{eqnarray} where
$a$,
$b$, $c$, $d$, $e$ are numbers representing Higgs meson vacuum values.
S(10),
A(120) and S(126) are the matrices of the Yukawa type constants which
couple the fermions to the 10, 120 and 126 Higgs mesons respectively; the
matrices S(10) and S(126) must be symmetric while A(120)  is
antisymmetric. Finally, $r\simeq 3$ is a renormalization factor for
comparing the quark masses with the charged lepton masses at a low energy
scale rather than at the grand unified scale; $s$ is a similar factor for
the neutrino masses. With the stated assumption that the $M_{L }$ term
dominates in Eq. (2.1) we get \begin{equation} {\rm Tr}(M_{\nu })\propto
{\rm Tr}(M_{d}) - r{\rm Tr}(M_{e})  \end{equation} which clearly also
holds when any number of 10's or 120's are present but only a single 126.
The matrices appearing in Eq. (2.3) are so far essentially unrestricted
complex ones. To proceed, we make the further assumption that the matrices
are hermitian. Then $ M_{d}$ and $M_{u} $ may each be brought to diagonal
form by unitary transformations. Thus the right hand side of Eq. (2.3) may
be evaluated to yield approximately, 
\begin{equation} {\rm
Tr}(M_{\nu})\propto m(b) -rm(\tau)\simeq 0 \end{equation} according to a
well known numerical success, based on the observation that $ r\simeq 3 $,
of grand unification \cite{RGE}. Note that we have not needed to assume
that the mass matrix has any zero elements.\footnote{In \cite{FrSt} a
similar mechanism was studied for $M_H$ where, in addition, a special
combined Fritzsch-Stech Ansatz was used. Here we are not making any
special Ansatz of this type for the mass matrices.}
 Even if the cancellation on
the right hand side of Eq. (2.4) is not perfect, it should still be a good
approximation. In an SO(10) model where the mass matrices are hermitian,
$M_{\nu }$ will be real symmetric. We will investigate this case and also
the possibility that the more general case holds.

 \section{Some notation} Our plan is to combine the Ansatz Eq. (1.1) with
experimentally obtained results on neutrino oscillations in order to learn
more about $M_{\nu}$ itself. For this purpose it may be helpful to set
down our notation \cite{notation} for the pieces of the effective ${\rm
SU}(2)_L\times {\rm U}(1)$ theory involving neutrinos and to make some
related remarks. 

    The free Lagrangian containing three two component massive fields is:
\begin{equation}
{\cal L}_{free}
    =-i\rho^{\dagger}\sigma_{\mu}\partial_{\mu}\rho -\frac{1}{2}(\rho^T
\sigma_2M_{\nu}\rho + h.c.), \label{Lfree} \end{equation} where $M_{\nu} =
M_{\nu}^T$ is the (not yet diagonalized) neutrino mass matrix of the
underlying theory to be identified with the matrix in Eq. (1.1). Note that
we are free to multiply the first mass term in Eq. (\ref{Lfree}) by an
overall arbitrary phase which is a matter of convention. It is
possible\cite{notation} to find a unitary matrix $U$ which brings $M_\nu$
to real, positive, diagonal form in the following way:  \begin{equation}
U^TM_\nu U={\hat M}, \quad {\hat M}={\rm diag}(m_1, m_2, m_3). 
\label{diagM} \end{equation} The mass diagonal fields $\nu$ are then
\begin{equation} \rho=U\nu.  \label{3.3} \end{equation} Similarly, the
column vector of left handed negatively charged leptons in the underlying
theory, $E_L$ is related to the mass diagonal fields $e_L$ by
\begin{equation} E_L = \Omega e_L, \label{3.4} \end{equation} where
$\Omega^{\dagger}=\Omega^{-1}$. 

Combining factors from Eq. (3.3) and Eq. (3.4) we obtain the unitary
mixing matrix, $K$
for the charged current weak interaction,
\begin{equation}
K = \Omega^{\dagger}U.
\label{3.5}
\end{equation}
This appears in the Lagrangian term,
\begin{equation}
{\cal L}_{int} = \frac{ig}{\sqrt 2}W_\mu^{-}{\bar e}_L\gamma_\mu K\nu
+h.c.,
\label{3.6}\end{equation}
where a conventional four component Dirac notation with $\gamma_5$
diagonal is
being employed and $\nu$ has only the first two components non zero. Next
we parameterize $K$\cite{notation}. It is possible to restrict ${\rm det}
K =1$
by adjusting an overall phase which can be absorbed in ${\bar e}_L$. Then
we write
\begin{equation}
K = \omega_0(\alpha)\omega_{23}(\eta_{23},\phi_{23})\omega_{12}(\eta_{12},
\phi_{12})\omega_{13}(\eta_{13},\phi_{13}) ,
\label{3.7}
\end{equation}
where
\begin{equation}
\omega_0(\alpha) = {\rm diag} (e^{i\alpha_1}, e^{i\alpha_2}, e^{i\alpha_3}) ,
\label{3.8}
\end{equation}
with $\alpha_3 =-(\alpha_1 + \alpha_2)$ and, for example
\begin{equation}
\omega_{12}(\eta_{12},\phi_{12})=
\left[\begin{array}{c c c}{\rm cos}\eta_{12}&e^{i\phi_{12}}{\rm
sin}\eta_{12}&0\\
-e^{-i\phi_{12}}{\rm sin}\eta_{12}&{\rm cos}\eta_{12}&0\\
0&0&1\end{array}\right] .
\label{3.9}
\end{equation}
Eq. (3.7) contains the eight parameters needed to characterize an arbitrary
unitary unimodular matrix. From the standpoint of Eq. (3.6) it can be
further simplified by using the freedom to rephase ${\bar e}_L\rightarrow {\bar e}_L \omega_0^{-1}(\alpha)$ without changing the free part of the
charged
lepton Lagrangian. On the other hand, the form of the mass terms in
Eq. (\ref{Lfree}) shows that the neutrino fields can not be rephased. Thus
a suitable minimal parameterization\footnote{This is written out in detail
in Eq (2) of \cite{GS}}
for $K$ in (3.6) is
\begin{equation}
K = \omega_{23}(\eta_{23},\phi_{23})\omega_{12}(\eta_{12},\phi_{12})
\omega_{13}(\eta_{13},\phi_{13}) ,
\label{3.10}
\end{equation}
involving three ``angles", $\eta_{ab}$ and three ``phases", $\phi_{ab}$.
Note the identity
\begin{equation}
\omega_0(\alpha)\omega_{ab}(\eta_{ab},\phi_{ab})\omega_0^{-1}(\alpha)
= \omega_{ab}(\eta_{ab},\alpha_a+\phi_{ab}-\alpha_b) .
\label{311}
\end{equation}
This identity may be used to transfer two of the phases $\phi_{ab}$
in Eq. (3.7) to a diagonal matrix on the right of $K$ as, for
example,
\begin{equation}
K =
\omega_{23}(\eta_{23},0)\omega_{12}(\eta_{12},0)\omega_{13}(\eta_{13},
\phi_{13})\omega_0^{-1}(\tau),
\label{3.12}
\end{equation}
where $\tau_1 + \tau_2 + \tau_3 =0$, which may be used instead of Eq. (3.10).

We also need the formula for the amplitude of neutrino oscillation.
For the case when a neutrino, produced by a charged lepton of type $a$,
``oscillates" to make at time $t$, a charged lepton of type $b$, we have
\begin{equation}
{\rm amp}(a\rightarrow b) = \sum_\alpha
K_{a\alpha}^*K_{b\alpha}e^{-iE_\alpha t} ,
\label{3.13}
\end{equation}
where the sum goes over the neutrinos of definite mass, $m_\alpha$.
Inserting the parameterization Eq. (3.12) into Eq. (3.13)
shows that the effect of the factor $\omega_0^{-1}(\tau)$ cancels out.
Thus for ordinary oscillations, $K$ is parameterized by three angles and
one CP violating phase as for the CKM quark mixing matrix. On the other
hand, the two additional CP violating phases $\tau_1$ and $\tau_2$
show up if one considers neutrino-antineutrino oscillations \cite{nan} or
neutrinoless double beta decay \cite{detail}.
  The formula for the probability, $P_{ab}$ is gotten by taking the
squared magnitude of Eq. (3.13) and replacing the exponential factor
$E_\alpha t$ by $(E+m_{\alpha}^2/(2E))L$, where $E$ is the neutrino energy
and $L$ is the oscillation distance. For practical reasons it is very
important to take account of the experimental uncertainties in $E$
and $L$. The simplest approximation\cite{Pdata} is to define $b=L/(4E)$
and assume that one can smear $P_{ab}$ with a Gaussian distribution in
$b$. $b_0$ is defined as the mean value and $\sigma_b$ as the standard
deviation appropriate to the particular physical setup. Then we find for
the smeared probability
\begin{eqnarray}
\left< P_{ab}\right> &=& \delta_{ab} -2\sum_{\alpha < \beta}[{\rm Re}(f_{\alpha\beta ab})
(1 - {\rm cos}(2b_0m^2_{\beta\alpha}){\rm exp}(-2\sigma_b^2(m^2_{\beta\alpha})^2))\\
\nonumber
&+&{\rm Im}(f_{\alpha\beta
ab}){\rm sin}(2b_0m^2_{\beta\alpha}){\rm exp}(-2\sigma_b^2(m^2_{\beta\alpha})^2)] ,
\label{3.14}
\end{eqnarray}
where $f_{\alpha\beta ab}=K_{a\alpha}K_{a\beta}^*K_{b\alpha}^*K_{b\beta}$
and $m^2_{\beta\alpha}=m^2_{\beta}-m^2_{\alpha}$. Notice that when
${\rm Im}(f_{\alpha\beta ab})=0$, $\left <P_{ab} \right>$ is independent of the sign of
$m^2_{\beta\alpha}$.

\section{Learning about $M_\nu$ from experiment}

Since ${\rm Tr} \left( M_\nu \right) = 0$ provides only two real
 equations for 12 real parameters, it is clear that it
 has a relatively small amount of predictivity.  In particular
 it cannot say much about the texture (e.g. possible zeroes) of $M_\nu$ which
 is suppposed to derive from a deeper theory than the standard model.
On the other hand, we shall see that our Ansatz is
 complementary to the results which should emerge
 from analysis of neutrino oscillation experiments.  Together,
 they should enable us to actually (with some conditions) reconstruct $M_\nu$.  

First, in this section, we shall consider $M_\nu$ to be
 hermitian so that the argument in favor of
 ${\rm Tr} \left( M_\nu \right) = 0$ presented in section II holds
 without any further assumptions.  Since $M_\nu$ is symmetric
 it must be real.  It can be brought to real
 diagonal form via a real rotation $R$ as ${R}^T M_\nu {R}$.
  However there is no guarantee that all eigenvalues
 of $M_\nu$ will be positive.  We can make them all
 positive by rephasing the diagonal fields with negative eigenvalues by 
a factor $i$ (see Eqs. (3.1) and (3.2)).  This means that
 the general diagonalizing matrix $U$ in Eq. (3.2) now takes the form 
\begin{equation}
U={R} \omega
\label{4.1},
\end{equation}
where $\omega_{ab} = \delta_{ab} \eta_b^{\frac{1}{2}}$ with
 $ \eta_b^{\frac{1}{2}}=1$ for a positive eigenvalue and
 $ \eta_b^{\frac{1}{2}}=i$ for a negative eigenvalue.
  We notice that Eq. (4.1) is of the form Eq. (3.12) for which
 we already noticed that the factor $\omega$ cancels
 out in the neutrino oscillation formula Eq. (3.14).  Furthermore only
 the square of the mass is relevant in Eq. (3.14).  Thus we choose
 to work in this section with some negative masses and no
 factor $\omega$ in Eq. (4.1).  

To avoid confusion, we remark that the factor $\omega$ in Eq. (4.1)
 does {\it not} introduce any CP violation in the theory 
\cite{wolf} since $M_\nu$ is real in any event.  

Now let us suppose that an experimental analysis of all
 neutrino oscillation experiments is made based on a formula
 like Eq. (3.14) (or one which treats the experimental
 uncertainties in a more sophisticated way).  Furthermore assume
 that the CP violating
 phase $\phi_{13}$ in Eq. (3.12) is negligible.  Then we should know
 the {\it magnitudes} of the squared neutrino mass differences
\begin{equation}
{\left( m_2 \right)}^2 - {\left( m_1 \right)}^2 = {A}, 
\quad {\left( m_3 \right)}^2 - {\left( m_2 \right)}^2 = {B},
\label{4.2}
\end{equation}
where $A$ and $B$ can be either positive or negative.  Then, assuming the
leptonic theory to be CP conserving, our Ansatz would imply
\begin{equation}
0 = {\rm Tr} \left( M_\nu \right) = {\rm Tr} \left( R {\hat M} R^T \right) = m_1 + m_2 + m_3,
\label{4.3}
\end{equation}
where Eq. (3.2) was used.  Eqs. (4.2) and (4.3) comprise three equations for the three neutrino masses $m_1$, $m_2$ and $m_3$.  We can solve to get:
\begin{eqnarray}
\left( m_1\right)^2 &=& \frac{2}{3} \left[ \sqrt{{A}^2 + {B}^2 + {AB}} -
 \left( {A} + \frac {B}{2} \right) \right], \\ \nonumber
m_2 &=& \frac{ {B} - {\left( m_1 \right)}^2}{2m_1}, \\ \nonumber
m_3 &=& - m_1 - m_2. 
\label{4.4}
\end{eqnarray}
This leads to a limited number of solutions, depending on sign choices. 

If we make the further assumption that the charged lepton
 mixing matrix $\Omega$ in Eq. (3.4) is approximately the unit
 matrix (this is expected to be a reasonable but not perfect approximation) we
 can identify $R \approx K$ which would be obtained from experiment.
  Then, using the masses found in Eq. (4.4), we could reconstruct $M_\nu$ as 
\begin{equation}
M_\nu \approx R {\hat M} R^T.
\label{4.5}
\end{equation}

To proceed, we need only insert the experimental results for $A$, $B$
 and $K$ in (4.4) and (4.5).  Of course, it is presumably the task of the next decade to solidify the experimental determination of these quantities.  We can, at the moment, only give a 
preliminary discussion.  For this purpose we will use the results of a recent preliminary analysis of all neutrino experiments by Ohlsson and Snellman \cite{Ohlsson}.  These authors found, by a least square analysis, a best fit for (our notation) 
$\left| 
A \right|$, $\left| B \right|$ and the leptonic mixing matrix $K$.  They used the formula Eq. (3.14) with a suitable choice of $b_0$ and $\sigma_b^2$ for each experiment.  Furthermore, they made the simplifying assumption that $K$ is real.  Finally they 
only 
searched for a fit in the range $10^{-4}\hskip 0.2cm {\rm eV}^2 \leq
 \left| {A} \right| \leq 10^{-3} \hskip 0.2cm {\rm eV}^2$,
 $0.2 \hskip 0.2cm {\rm eV}^2 \leq \left| {B} \right| \leq 2
 \hskip 0.2cm {\rm eV}^2$.  This range corresponds to mass difference
 choices for which the MSW effect \cite{MSW} for solar 
and atmospheric neutrinos is not expected to be important
 and so greatly simplifies the analysis.  Thus there is no
 guarantee that the solution of \cite{Ohlsson} is unique.
  Altogether they fit sixteen different solar neutrino,
 atmospheric neutrino, accelerator 
and reactor experiments, including LSND.  The best fit is: 
\begin{equation}
\left| {A} \right| = 2.87 \times 10^{-4} \hskip 0.2cm {\rm eV}^2, 
\quad\left| {B}
\right| = 1.11 \hskip 0.2cm {\rm eV}^2,
\label{4.6}
\end{equation}
for the squared mass differences and 
\begin{equation}
K_{exp} = \left[
\begin{array}{c c c}
0.7052&0.7052&0.0732\\
-0.6441&0.5940&0.4820\\
0.2964&-0.3871&0.8731
\end{array}
\right]
\label{4.7}
\end{equation}
for the lepton mixing matrix $K$.  As discussed above we will identify $K \approx R$ here, keeping $K$ real but allowing for negative masses.  The best fit matrix K was obtained to be similar but not identical to the 
``bimaximal mixing'' matrix \cite{Barger}.  

With the best fit squared mass differences in Eq. (4.6), our model predicts,
 from the first of Eq. (4.4), eight different possibilities.  These
correspond  to four different sign configurations for $A$ and $B$ times
 the two possible signs for $m_1$.  However, only 
two of these eight are essentially different;  these are
\begin{eqnarray}
&{\rm type}&\hskip 0.2cm {\rm I}: \quad m_1= 0.6082 \hskip 0.2cm {\rm
eV}, \quad m_2 = 0.6084 \hskip 0.2cm {\rm eV}, \quad m_3 = -1.2166
\hskip 0.2cm {\rm eV} \\
&{\rm type}& \hskip 0.2cm {\rm II}:\quad  m_1 = 1.053701 \hskip 0.2cm
{\rm eV}, \quad m_2 = -1.053565 \hskip 0.2cm {\rm
eV}, \quad  m_3 = -0.000136 \hskip 0.2cm {\rm eV}
\label{4.8}
\end{eqnarray}
The other solutions correspond to interchanging whichever of $\left| m_1
\right|$ and $\left| m_2 \right| $ is greater (which has only
 a negligible effect since they are almost degenerate) or reversing  the
signs of all masses.  Physically it is clear what 
is happening: the smallness of $\left| {A} \right|$ compared to
 $\left| {B} \right|$ in Eq. (4.2) forces $\left| m_1 \right| \approx 
\left| m_2 \right|$.  Then we have either $m_1 \approx m_2$ with, using
the constraint Eq. (4.3), $m_3 \approx -2m_1$
 or $m_1 \approx -m_2$ with $m_3$ very small.  

Since we have assumed the neutrinos to be of Majorana type for our
plausibility argument in section II, their interactions will violate
lepton number.  Then they should mediate neutrinoless double beta
decay $\left( \beta \beta _{0\nu} \right)$ \cite{bilgiungrim}.  Such a process has not yet been observed and an upper bound has been set for the relevant quantity 
\begin{equation}
\left< m_\nu \right> \equiv \left| {\sum_{\alpha = 1}^3}
{\left(K_{1\alpha}
\right)}^2 m_\alpha \right|
.
\label{4.10}
\end{equation}
The best upper bound at present is \cite{Heidelberg} $\left< m_\nu \right> \leq
0.2-0.6$ eV, reflecting some uncertainty in the estimation of the needed nuclear matrix elements.  

Substituting the best fit for the matrix K from Eq. (4.7) together
with our results in Eqs. (4.8) and (4.9) into Eq. (4.10) yields predictions for the two cases:
\begin{eqnarray}
{\rm type} \hskip 0.2cm {\rm I}: \quad \left< m_\nu \right> &=& 0.60
\hskip 0.2cm {\rm
eV}, \\ 
{\rm type} \hskip 0.2cm {\rm II}: \quad \left< m_\nu \right> &=& 6.7
\times 10^{-5} \hskip 0.2cm  {\rm eV}.
\label{4.11}
\end{eqnarray}
Both solutions seem to be acceptable, the type I case marginally but
 the type II case definitely.  Note that the small value for $\left< m_\nu 
\right>$ in the type II case is due to the best fit prediction
\cite{Ohlsson}
 $K_{11} = K_{12}$ and also to the fact that
 $m_2$ is negative.  The same value would clearly result if we made $m_2$ positive and set $K_{12} = 0.7052i$ as discussed around Eq. (4.1) above.

Finally, let us reconstruct the underlying neutrino mass matrices for each of the two cases.  We use Eq. (4.5) based on the assumption that $M_\nu$ is real and also our ansatz to find (in units of eV):
\begin{eqnarray}
{\rm type}\hskip 0.2cm {\rm I}: \quad  M_\nu &=& M_\nu^T \approx \left[ \begin{array}{c c c}
0.5985&-0.0643&-0.1167\\
-0.0643&0.1843&-0.7680\\
-0.1167&-0.7680&-0.7828 
\end{array}
\right], \\
{\rm type} \hskip 0.2cm {\rm II}: \quad M_\nu &=& M_\nu^T \approx \left[ \begin{array} {c c c}
6.7 \times 10^{-5}& -0.9199&0.5078\\
-0.9199&0.0654&0.0410\\
0.5078&0.0410&-0.0654\\
\end{array}
\right].
\label{4.13}
\end{eqnarray}
The type I matrix does not have an excellent candidate
 for a ``texture'' zero.  However the small value
 of ${\left( M_\nu \right)}_{11}$ in the type II case is
 certainly suggestive.  These matrices lead to neutrino
 masses and a mixing matrix which give a 
best fit to all present data.  It will be interesting to see if either of them hold up in the future.  

Incidentally, on comparing Eqs. (4.13) and (4.14) it is amusing to
observe the large difference in two mass matrices ``generated'' in the
 same way except with respect to how ${\rm Tr} \left( M_\nu \right)=0 $ is
satisfied.

\section{Case of complex $M_\nu$}

It seems interesting to also investigate the Ansatz ${\rm Tr} (M_\nu) =0$
when $M_\nu$ is no longer restricted to be real.  This also raises the
problem of constructing the unitary diagonalizing matrix $U$ in Eq.(3.2),
in terms of the experimentally measured lepton mixing matrix $K_{exp}$.
For simplicity, as before, we will make the approximation that the
charged lepton diagonalizing matrix $\Omega$ is the unit matrix.

Apart from an overall (conventional) phase we may write
\begin{equation}
U = \omega_0(\sigma) K_{exp} \omega_0^{-1}(\tau),
\end{equation}
where the 2-parameter quantity $\omega_0$ was defined in Eq. (3.8).  Since
$K_{exp}$ has four parameters $U$ in Eq. (5.1) is described by eight
parameters.   As mentioned before, the two parameters in
$\omega_0^{-1}(\tau)$ are not measurable in neutrino oscillation
experiments but show up when one considers $(\beta\beta)_{o\nu}$.
The two parameters in $\omega_0(\sigma)$ may be eliminated, for
experimental purposes, by rephasing the charged leptons.
 However, for the theoretical purpose of reconstructing  the underlying
neutrino mass
matrix $M_\nu$, their existence cannot be ruled out.   (They also do not
contribute to $(\beta\beta)_{0\nu}$.)   

For the purpose of relating the Ansatz on $M_\nu$  to the physical
neutrino masses in ${\hat M}$, we note

\begin{equation}
{\rm Tr} (M_\nu) =
{\rm Tr} (
K_{exp}^{-1} \omega_0^{-1} (2\sigma) K^*_{exp}\omega_0(2\tau)
{\hat M} ).
\end{equation}

\begin{figure}
\centering
\epsfig{file=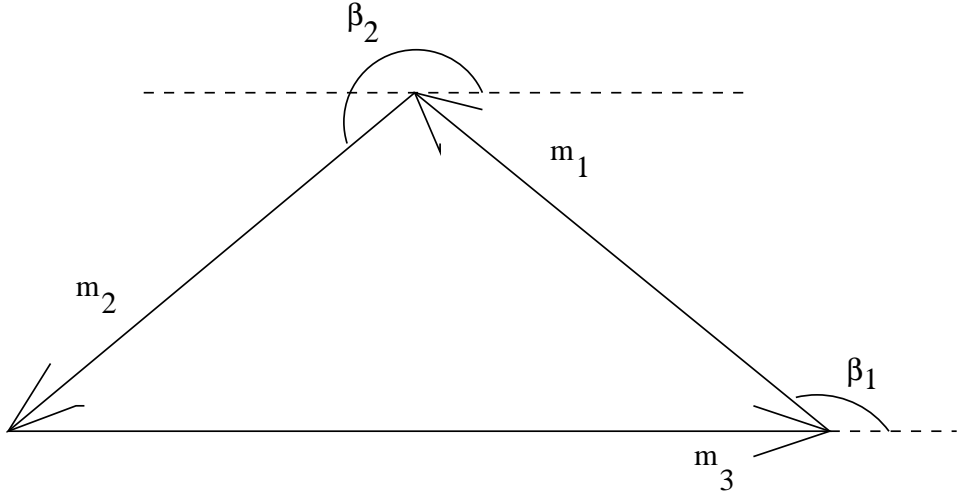, height=5in, angle=270}
\caption
{Geometrical picture of Eq. (5.4).}
\label{trianglefig}
\end{figure}

For further simplicity of the analysis we adopt the special case
$\omega_0(2\sigma)=1$ and also identify $K_{exp}$ with the real best fit
in Eq. (4.7); our Ansatz now reads

\begin{equation}
0={\rm Tr}(M_\nu)={\rm Tr} \left(\omega_0(2\tau) {\hat M} \right).
\end{equation} 
With the redefinitions $\beta_1=4\tau_1 + 2\tau_2$ and $\beta_2 = 2 \tau_1
+ 4 \tau_2$, Eq. (5.3) becomes
\begin{equation}
e^{i\beta_1} m_1 + e^{i\beta_2} m_2 + m_3 = 0
\end{equation} 
This may be conveniently visualized as the vector triangle shown in Fig.
\ref{trianglefig}. 

Combining Eqs. (5.4) and (4.2) gives four real equations for the five
unknown quantities ($m_1,m_2,m_3,\beta_1,\beta_2$).  Thus we have 
(for each
set of ($A,B$) sign choices) a one parameter family of solutions.  It is
convenient to choose this parameter to be $m_3$.  Then $m_1$ and $m_2$
may be found from the equations (4.2), provided that solutions exist.
In this way all three sides of the triangle in Fig.\ref{trianglefig} are determined.  
The angles may finally be found as 
\begin{eqnarray}
{\rm cos} \beta_2 &=& {{m_1^2 - m_2^2 -m_3^2}\over {2 m_2m_3}}\nonumber
\\
{\rm sin} \beta_1 &=& -{ m_2\over m_1} {\rm sin} \beta_2.
\end{eqnarray}

We also need to investigate the constraint arising from the
non-observation of $(\beta\beta)_{0\nu}$.
Eq. (4.10) now becomes, with Eq. (5.1) as the mixing matrix
\begin{equation}
\langle m_\nu \rangle =
\left| 
m_1 (K_{exp\hskip .2cm 11})^2 e^{-i\beta_1} +
m_2 (K_{exp\hskip .2cm 12})^2 e^{-i\beta_2} +
m_3 (K_{exp\hskip .2cm 13})^2 
\right|.
\end{equation}

Using the Ansatz constraint Eq. (5.4), Eq. (5.6) may be rewritten as
\begin{equation}
\langle m_\nu \rangle =
\left| 
\left[
(K_{exp\hskip .2cm 12})^2 - (K_{exp\hskip .2cm 11})^2
\right]
m_2 e^{-i\beta_2} +
\left[
(K_{exp\hskip .2cm 13})^2 - (K_{exp\hskip .2cm 11})^2 
\right]
m_3
\right|.
\end{equation}
This form is very convenient when identifying $K_{exp}$ with the best fit
solution in Eq. (4.7).  In the present context such an identification
corresponds to CP violation for the $(\beta\beta)_{0\nu}$ process but not 
for usual neutrino oscillations.  Since the (11) and (12)
matrix elements are equal in Eq. (4.7) we find the simple result
\begin{equation}
\langle m_\nu \rangle =
0.49 m_3.
\end{equation}
Thus if the upper bound on $\langle m_\nu \rangle$ is conservatively
identified as in the $0.2 - 0.6$ eV range, we should have in this case
\begin{equation}
m_3 \le 0.41 - 1.22 {\rm eV}.
\end{equation}
In the present complex case there is a continuum of possible 
solutions labelled by those values of $m_3$ satisfying Eq. (5.9),
rather than just the two possibilities found in Eq. (4.8) and Eq. (4.9).
Actually, the continuum separates roughly into two classes similar to
either Eq. (4.8) or Eq. (4.9).   In the generalized type I class, $m_3$ is
of the 
order $B = 1.11$ eV while $m_1$ and $m_2$ are related to nearly
oppositely directed vectors in 
Fig. \ref{trianglefig} and are also of the order $B$.
    In the generalized type 
II class, $B$ is negative; $m_1$ and $m_2$ correspond to vectors of order
$|B|$ which are  
oppositely directed to each other, while $m_3$ ranges from very small
to order $\left| B \right|$.

Given the bound Eq. (5.9) from the non-observation of
$(\beta\beta)_{o\nu}$, there are important limitations on the allowed
$m_3$ values for type I solutions.  In this case $B$ is positive so the
equation
\begin{equation}
(m_2)^2 = (m_3)^2 - B = (m_3)^2 -(1.11 \hskip 0.2cm {\rm eV})^2,
\end{equation}
will only allow solutions for $m_3 > 1.11$ eV.  This range is 
barely compatible with Eq. (5.9).  Thus the type II case where $B< 0$ and
$m_1 \approx m_2 > m_3$ seems most probable.

As an example of a solution for complex $M_\nu$, consider choosing
$m_3=0.2$ eV.  Then a solution is obtained with (compare with Fig. \ref{trianglefig})

\begin{equation}
m_1\approx m_2 \approx 1.128\hskip .2cm {\rm eV}, 
m_3 = 0.2 \hskip .2cm{\rm eV},
\beta_1 \approx 95.1^o,\beta_2 \approx 264.9 ^o.
\end{equation}
The matrix $U$, which diagonalizes $M_\nu$ is obtained from Eq. (5.1), with
the approximation $\omega_0 \left( \sigma \right) = 1$, and with now:
\begin{equation}
\omega_0^{-1}(\tau)= {\rm diag}
(0.976 + 0.216 i, 0.301 - 0.953 i, 0.500 + 0.866 i)
\end{equation}
This factor introduces CP violation in the $(\beta\beta)_{0\nu}$ process
but not in ordinary neutrino oscillations.  

Finally the underlying neutrino mass matrix, $U{\hat M}U^T$
is ``reconstructed'' as (in units of eV):
\begin{equation}
M_\nu= M_\nu^T =
\left[
\begin{array}{c c c}
0.049 - 0.085 i & -0.855 - 0.481 i & 0.459 + 0.287 i\\
                & 0.076 + 0.009 i &-0.025 + 0.131 i \\
                &                  & -0.125 + 0.077 i
\end{array}
\right].
\end{equation}
This is structurally similar to the real type II solution displayed in
Eq. (4.14), although the suppression of the (11)
element is not so pronounced.  Notice that $m_3$ is considerably smaller
than the almost degenerate pair $m_1$ and $m_2$.   Furthermore $m_1$ and
$m_2$ are large enough to possibly play some role in astrophysics.

\section{Summary and Discussion}

We investigated the Ansatz ${\rm Tr} \left( M_\nu \right) = 0$ for the
underlying (pre-diagonal) three generation neutrino mass matrix.  It was
motivated by noting that in an SO(10) grand unified model where $M_\nu$ was
taken to be real (CP conserving), it corresponds to the well known
unification of b quark and $\tau$ lepton masses.  While not very predictive
by itself it yields information complementary to what would be gotten from
a complete three flavor analysis of all lepton number conserving neutrino
oscillation experiments.  Specifically from the specification of the
magnitudes of two neutrino squared mass differences and also of the
leptonic mixing matrix we can, with some assumptions, find the neutrino
masses themselves and ``reconstruct'' $M_\nu$.  This determination can be
sharpened by consideration of the constraints imposed by non-observation of
neutrinoless double beta decay.  

For the purpose of testing our Ansatz we employed the results of a
reasonable best fit to all present neutrino experiments (including LSND) by
Ohlsson and Snellmann \cite{Ohlsson}.  This fit will inevitably be
improved in the
next few years as new experiments are completed.  They were able to fit the
data without assuming any CP violation.  This agrees with assuming $M_\nu$
to be real.  We found two essentially different solutions in that case.
The first features two neutrinos having approximately equal mass 0.608 eV
and a third neutrino of mass 1.217 eV.  This solution is on the borderline
of being ruled out by non-observation of $\left( {\beta \beta}_{0\nu}
\right)$.  The second solution has two neutrinos with approximately
degenerate mass 1.054 eV and a third neutrino with a mass $1.36 \times
{10}^{-4}$ eV.  This solution is very safe from being ruled out by $\left(
{\beta \beta}_{0\nu} \right)$ experiments.  It also features a
reconstructed  $M_\nu$ which has an extremely small (11) element.
Note that, for both solutions, even though $M_\nu$ is real there are some
(CP conserving) factors of $i$ in the mixing matrix when all masses are
taken to be positive.  Alternatively one may have no $i$'s in the mixing
matrix while allowing some masses to be negative.   The latter form is
useful for seeing intuitively how ${\rm Tr}(M_\nu)$=0 is possible.

The case of matching the above best fit data to a complex $M_\nu$ was also
considered.  In this situation there are CP violating phases in the lepton
mixing matrix which affect the $(\beta\beta)_{0\nu}$ process but do not
affect ordinary total lepton number conserving neutrino oscillations.
Such
phases could also be measurable in principle with the observation of a
decay like $\tau^- \rightarrow \pi^-\pi^-\mu^+$.  The case of complex
$M_\nu$ allows a larger number of solutions.  With a simplifying
assumption there is a one parameter family of allowed neutrino mass
sets. Roughly, these fall into one of the two types already
encountered for real $M_\nu$.

A question of some interest is whether the neutrinos are massive enough to
play a role in cosmology.   The relevant criterion \cite{cosmos} for this
to occur is
usually stated as ${\sum}_{a} m_a > 6$ eV.   For the type II solutions with
complex $M_\nu$ we have found the largest mass sum to be about 4.5 eV
corresponding to $m_1\approx m_2 = 1.62$ eV and $m_3\approx 1.22$ eV.
However this is on the very border of acceptability for non observation of
$(\beta\beta)_{0\nu}$.

Future best fits to the neutrino oscillation data can easily be
accommodated in the present framework.  Of course, the predictions
for neutrino masses and mixings will
depend on this input.

\acknowledgments
We would like to thank H. Benaoum for useful comments on the manuscript. 
This work has been 
supported in part by the US DOE under contract DE-FG-02-85ER 40231.

\end{document}